\documentclass[12pt,preprint]{aastex}
\begin{document}

\title{Interferometric Observations of Powerful CO Emission from three
 Submillimeter Galaxies at $z=2.39$, $2.51$ and $3.35$ 
\footnote{Based on observations obtained at the IRAM Plateau de Bure
Interferometer. IRAM is funded by the Centre Nationale de la Recherche
Scientifique (France), the Max-Planck Gesellschaft (Germany), and the
Instituto Geografico Nacional (Spain).}}

\author{R.Neri\altaffilmark{1}; R.Genzel\altaffilmark{2,3}; 
R.J.Ivison\altaffilmark{4}; F.Bertoldi\altaffilmark{5}; 
A.W.Blain\altaffilmark{6}; S.C.Chapman\altaffilmark{6}; 
P.Cox\altaffilmark{7}; T.R.Greve\altaffilmark{8}; A.Omont\altaffilmark{9}; 
D.T.Frayer\altaffilmark{10}}

\altaffiltext{1}{Institut de Radio Astronomie Millim\'etrique (IRAM),
St.Martin d'H\`{e}res, France\ \ (neri@iram.fr)}
\altaffiltext{2}{Max-Planck Institut f\"{u}r extraterrestrische Physik
(MPE), Garching, FRG\ \ (genzel@mpe.mpg.de)}
\altaffiltext{3}{Department of Physics, University of California,
Berkeley, USA}
\altaffiltext{4}{Astronomy Technology Centre, Royal Observatory,
Edinburgh, UK\ \ (rji@roe.ac.uk)}
\altaffiltext{5}{Max-Planck Institut f\"{u}r Radioastronomie (MPIfR),
Bonn, FRG\ \ (bertoldi@mpifr-bonn.mpg.de)}
\altaffiltext{6}{Astronomy Department, California Institute of
Technology, Pasadena, USA\ \ (awb@astro.caltech.edu, schapman@irastro.caltech.edu)}
\altaffiltext{7}{Institut d'Astrophysique Spatiale, Universit\'{e} de
Paris Sud, Orsay, France\ \ (cox@iap.fr)}
\altaffiltext{8}{Institute for Astronomy, University of Edinburgh,
Edinburgh, UK\ \ (tgreve@roe.ac.uk)}
\altaffiltext{9}{Institut d'Astrophysique Spatiale, CNRS \& Universit\'{e} 
de Paris 6, Paris, France\ \ (omont@iap.fr)}
\altaffiltext{10}{SIRTF Science Center, California Institute of
Technology, Pasadena, USA\ \ (frayer@ipac.caltech.edu)}
%\author {I.Smail }
%\affil{
%Institute for Computational Cosmology, University of Durham, Durham, UK
%}

\clearpage
\begin{abstract}
We report IRAM millimeter interferometry of three $z\sim$2.4 to 3.4,
SCUBA deep field galaxies. Our CO line observations confirm the
rest-frame UV/optical redshifts, thus more than doubling the number of
confirmed, published redshifts of the faint submm-population and
proving their high-$z$ nature. In all three sources our measurements
of the intrinsic gas and dynamical mass are large ($10^{10}$ to
$10^{11}$ M$_{\odot}$). In at least two cases the data show that the
submm sources are part of an interacting system. Together with recent
information gathered in the X-ray, optical and radio bands our
observations support the interpretation that the submm-population, at
least the radio detected ones, consists of gas rich (gas to dynamical
mass ratio $\sim$0.5) and massive, interacting starburst/AGN systems.
\end{abstract}

\keywords{cosmology: observations - galaxies: formation -galaxies:
high-redshift - galaxies: evolution}

\section{Introduction}

The extragalactic far-IR/submm background is probably
dominated by luminous infrared galaxies (LIRGs/ULIRGs:
L$_{\mathrm{IR}}\sim 10^{11.5}$ to 10$^{13}$ L$_{\odot}$) at $z\ga1$
(e.g.\thinspace\ Smail et al.\ 1997; Bertoldi et al.\ 2002; Scott et
al.\ 2002; Cowie et al.\ 2002). However, far-IR/submm sources
in most cases have relatively poorly known positions and frequently
have only weak counterparts in the rest-frame UV and optical (Smail et
al.\ 2000, 2002; Dannerbauer et al.\ 2002). As a result redshifts have
thus far been confirmed with CO interferometry for only two of the
$\sim$\,100 detected systems (Blain et al.\ 2002). Recently a subgroup
of the authors have obtained optical spectroscopic redshifts for a
number of sources detected with the SCUBA camera at
850\textrm{$\mu$m}, to a large extent aided by more precise positions
derived from deep 1.4\,GHz VLA observations of the same fields. Here we
report the first results on the millimeter CO line follow-up of these
submm sources. We believe these observations mark a sensitive
breakthrough in the notoriously difficult study of the faint
far-IR/submm galaxy population.

\section{Observations}

The observations were carried out between late summer 2002 and winter
2003 with the IRAM Plateau de Bure interferometer, consisting
of six 15m-diameter telescopes. We used the compact D configuration,
and for follow-up observations of two of the sources the more extended
BC configurations. The correlator was configured for CO line and
continuum observations to simultaneously cover 580 MHz in the 3\,mm
and 1.3\,mm bands. The frequency settings were adjusted for all three
sources to optimize the CO line centering in the
bandpass. SMMJ04431+0210 was observed between Sep '02 and Feb '03 in D
and BC configurations for a total integration time of 22
hrs. SMMJ09431+4700 was observed in D configuration only, in Nov '02,
for 13\,hrs.  SMMJ16368+4057 was observed between Sep '02 and Feb '03
for 24\,hrs.  All sources were observed in very good observing
conditions. We calibrated the data using the CLIC program in the IRAM
GILDAS package. Passband calibration used
one or more bright quasars. Phase and amplitude variations within each
track were calibrated out by interleaving reference observations of
nearby quasars every 20 minutes. The overall flux scale for each epoch
was set on MWC 349. After flagging bad and high phase noise data (less
than 2\% at 3\,mm), we created data cubes in natural weighting.

\section{Results}

Figures 1 and 2 show the CO spectra and maps. The derived properties
are listed in Tables 1 and 2. We adopt a flat, $\Omega_{\mathrm{M}}$ =
0.3 $\Lambda $-cosmology with $H_0$=70 km\thinspace
s$^{-1}$Mpc$^{-1}$. To convert CO luminosities to gas masses,
including a 37\% correction for helium, we adopt, under the assumption
of constant brightness temperature for the lowest rotational
transitions from 1-0 to 4-3, a factor $\alpha $=0.8 M$_{\odot }$/(K
km\thinspace s$^{-1}$ pc$^{2}$)=0.2$\alpha $ (Galactic), as derived
from observations of $z$\,$\sim$\,0.1 ULIRGs (Downes \& Solomon
1998). The gas masses are probably uncertain by a factor of at least
2. We estimate dynamical masses from M$_{\rm dyn}$ $\sin^{2}i$
(M$_{\odot})=4\times10^{4} \Delta v_{\rm FWHM}^{2}R$.
% 
%
%\qquad \qquad \qquad  M$_{\rm dyn}$ (sin $i$)$ ^{2}=4.0\times 10^{4}\Delta
%v_{\rm FWHM}^{2}R$\ \ \ \ (M$_{\odot }$)\ . \ \ \ \ \ \ \ \ \ \ \ \ \ \ (1)
%
%
Here we assume that the gas emission comes from a rotating disk of
outer radius $R$ (kpc) observed at inclination angle $i$. In a merger
model the dynamical masses would be a factor of 2 larger (Genzel et
al.\ 2003). The numerical constant incorporates a factor of 2.4
between observed FWHM velocity width of the line emission, $\Delta
v_{\rm FWHM}$ (km/s), and the product of rotation velocity and sin$i$.
This factor is estimated from model disks taking into account local
line broadening, beam and spectral smearing. We deduce IR luminosities
from the 850$\mu$m continuum flux densities S by adopting a modified
greybody model (T=40 K) with $\varpropto\nu^{1.5}$ dependent
emissivity, such that in the range from $z$=2 to 3.5 L$_{\rm
IR}$ (L$_{\odot}$) $=1.9\times 10^{12}\,$S$_{850}$
%
% \ \ \ \ \ \ \ \ \ \ \ \ \ \ \ \ \ \ \ \qquad L$_{\rm IR}$ (L$_{\odot
% }$) $=1.9\times 10^{12}\,$S$_{850}\ ,\qquad \qquad $(2)
%
%
with S$_{850}$ in mJy (Blain et al.\ 2002). These luminosities are
uncertain by a factor of 2 to 3 since dust temperature and emissivity
law may vary from source to source (Blain et al.\ 2003). In the
following, all linear sizes, masses and luminosities are corrected for
the foreground lensing factors in Table 2.

%\subsection{SMMJ04431+0210 ($z=2.51$)}
%
\noindent 
{\em SMMJ04431+0210}\/ ($z$=2.51) was originally found in the SCUBA
Lens Survey (S$_{850}$=7.2\thinspace mJy; Smail et al.\ 1997; 2002).
It is located behind the $z$=0.18 cluster MS0440+02. Smail et al.\
(1999) identified the submm source with the $K=19.4$ extremely red
object (ERO) N4 about 3$^{\prime\prime}$ NW of an edge-on cluster
spiral galaxy N1 (N4: $R$--$K$=6.3; Frayer et al.\ 2003).  Frayer et
al.\ deduced a redshift of $z$=2.5092$\pm $0.0008 from
H$\alpha$/[N\thinspace \textsc{ii}]/[O\thinspace \textsc{iii}] line
emission. The rest frame optical line ratios suggest that N4 is a
composite starburst/narrow line AGN. We adopt the foreground lens
magnification of 4.4 deduced by Smail et al.\ (1999). Our BCD
configuration data show a strong CO 3-2 line centered at $z$=2.5094$
\pm $0.0002, $+$17 ($\pm$\,17) km\thinspace s$^{-1}$ redward of the
nominal redshift of the H$\alpha $ line. The line width is FWHM 350
$\pm $ 60 km\thinspace s$^{-1}$, somewhat smaller than that of
H$\alpha $ (520 km\thinspace s$^{-1}$). The integrated CO line flux
corresponds to a total gas mass of 8$\times $10$^{9}$ M$_{\odot
}$. Most of the CO line emission comes from within 1$^{\prime \prime
}$. The CO emission centroid is 1.1$^{\prime \prime }$ SW of
the near-IR position of N4, as determined by a new astrometric
solution of the near-IR/radio astrometry we obtained by comparing USNO
stars with radio sources in the field (N4:
04$^{\mathrm{h}}$43$^{\mathrm{m}}$07.25$^{\mathrm{s}}$ 02$^{\circ
}$10$^{\prime }$24.4$^{\prime \prime }$ (J2000); the uncertainty is
$\pm $0.5$^{\prime \prime }$). This new position of N4 is 2.2$^{\prime
\prime }$ E and 0.6$^{\prime \prime }$ S of the position
reported by Smail et al.\ (1999) based on the APM coordinate
system. The 1.3\thinspace mm continuum data show a marginally
significant detection (1.1\,mJy, 3.7$\sigma $) near the position of
N4. We set a limit to the CO 7-6 emission of $\leq $ 0.8 Jy
km\thinspace s$^{-1}$ (2$\sigma $). For comparison, the H$\alpha $
emission exhibits a velocity gradient of $\geq 400$ km\thinspace
s$^{-1}$ over about 1$^{\prime\prime}$ and along the slit at p.a.\
$-14^{\circ }$, that is, at about 50$^{\circ }$ relative to the
direction of the extended CO emission. The centroid of the H$\alpha $
emission is on N4. It thus appears that the rest-frame submm
and optical observations sample a similar region (size $\sim
$1$^{\prime \prime }$) but with some differences in the spatial
structure in the two wavelength ranges.  SMMJ04431+0210 has by far the
lowest intrinsic IR luminosity ($3\times 10^{12}\,$L$_{\odot }$)
and gas/dynamic mass of our three galaxies. We deduce an upper limit
to the dynamical mass of 4.5$\times 10^{9}\sin^{-2}i$\,M$_{\odot }$
for a source diameter of $\le 1^{\prime \prime }$. In terms of
luminosity, gas and dynamical mass, SMMJ04431+0210 thus resembles
local ULIRGs.

%\subsection{SMMJ09431+4700 ($z=3.35$)}
%
\noindent 
{\em SMMJ09431+4700}\/ ($z$=3.35) was first identified by Cowie et
al.\ (2002) in a deep SCUBA map of the $z$=0.41 cluster Abell 851
(S$_{850}$=10.5 mJy). They estimated a foreground lens magnification
of 1.2. Ledlow et al.\ (2002) proposed that the counterpart of the
SCUBA source is the 1.4 GHz radio source H6 (72$\mu $Jy), for which
they identified a redshift of $z$=3.349 from Ly$\alpha $. H6 appears
to be a UV bright, narrow line Seyfert 1 galaxy. We find strong CO 4-3
emission with a flat-topped profile and FWHM 420 km\thinspace s$^{-1}$
centered at $z$=3.3460 $\pm $ 0.0001, $-$207 ($\pm $7) km\thinspace
s$^{-1}$ blueward of the nominal Ly$\alpha$ redshift. The
1.3\thinspace mm continuum was also detected with 2.3 $\pm $
0.4\thinspace mJy and is centered at the same position. CO line and
continuum emission are centered 3.8$^{\prime \prime }$ W and
1$^{\prime \prime }$ S of the position of H6 (24 kpc in the source
plane), positionally coincident with the second, weaker 1.4 GHz source
H7 (55 $\mu $Jy) at
09$^{\mathrm{h}}$43$^{\mathrm{m}}$03.7$^{\mathrm{s}}$ 47$^{\circ
}$00$^{\prime }$15.1$^{\prime \prime }$ (J2000) (Fig.2).  H6 and H7
are very probably physically related. The gas mass deduced from the CO
4-3 flux is 2.1$\times$10$^{10}$ M$_{\odot }$, and the IR luminosity
based on S$_{\rm 850}$ is 1.7$\times $10$^{13}$\,L$_{\odot }$. For an
assumed source size of 1$^{\prime \prime }$ of H7 we infer a dynamical
mass of 2.5$\times$10$^{10} \sin^{-2}(i)$ M$_{\odot }$. A lower limit
to the virial mass of the H6/7 system is 6$\times$10$^{10}$
M$_{\odot}$.

%\subsection{SMMJ16368+4057 ($z=2.39$)}
%
\noindent
{\em SMMJ16368+4057}\/ ($z$=2.39, Elais\,N2\,850.4) was identified by
Ivison et al.\,(2002; S$ _{850}$= 8.2\,mJy) in the 8mJy SCUBA blank
field survey of the Elais N2 field (Scott et al.\ 2002) with a
220\textrm{$\mu $Jy} bright 1.4 GHz radio source. Optical spectroscopy
by Chapman et al.\ (2003) and Smail et al.\ (2003) showed bright
Ly$\alpha $, N\thinspace \textsc{v}, C\thinspace \textsc{iv},
[O\thinspace \textsc{ii}] and [O\thinspace \textsc{iii}] emission with
a complex spatial and velocity structure. Smail et al.\ (2003)
proposed that N2\,850.4 consists of a UV bright starburst galaxy at
$z$=2.380 ($\pm $0.002), plus a Seyfert 2 galaxy at $z$=2.384 ($\pm
$0.003). There is no evidence for gravitational lensing. We also
detected the 1.3\thinspace mm continuum emission (2.5 $\pm $ 0.4 mJy)
and CO 7-6 emission. The CO 3-2 emission is very broad (840
km\thinspace s$^{-1}$ FWHM) and is centered at $z$=2.3853 ($\pm
$0.0004), $+$115 km\thinspace s$^{-1}$ ($\pm $36 km\thinspace
s$^{-1}$) redward of the nominal redshift of the Sey2 nucleus. CO 7-6
emission is tentatively detected in a narrow component centered
$-$200\thinspace km\thinspace s$ ^{-1}$ of the CO 3-2 line
centroid. At the 7-6 peak the observed 7-6/3-2 brightness temperature
ratio is 0.5 $\pm $ 0.2.  Large velocity gradient modeling of this
ratio indicates that the higher excitation CO 7-6 emission may come
from a specific warm (T $\geq $ 50 K) and dense (n(H$_{2})\geq
10^{4}$\thinspace cm$^{-3}$) region. The absolute astrometry of the
rest-frame optical/UV, submm and radio positions (each $\pm $0.3$
^{\prime \prime }$) is not yet sufficient to establish with certainty
the relative locations of the emission sources at different
wavebands. Relative positions are more precise and indicate that the
UV bright source is about 0.5$^{\prime \prime }$ W or SW of the
optical source, and both have a size of about 0.7$^{\prime \prime }$,
or 5.7\thinspace kpc.  Likewise, we find that the different submm
components are spread over $\sim $ 0.7$^{\prime \prime }$, with the CO
7-6 and submm continuum about 0.3$ ^{\prime \prime }$ to the SW, while
the CO 3-2 emission is centered to the NE. Keeping in mind that these
differences are marginally significant we note that the spatial
offsets appear to be along p.a.\ 45$^{\circ }$, the direction of the
separation between the UV and optical line emission sources (Smail et
al.\ 2003). The 1.3 mm continuum ($1.6\times 10^{13}\,$L$_{\odot }$)
is unresolved, with an upper limit of about 1$^{\prime \prime }$. The
gas mass estimated from the CO emission is about $5.4\times 10^{10}$
M$_{\odot }$, and the dynamical mass is 4.5$\times 10^{10}\sin^{-2}i$
M$_{\odot }$ for an adopted source diameter of 0.7$^{\prime \prime }$.

\section{Discussion}

Our observations of three SCUBA selected galaxies confirm the
redshifts identified from rest-frame UV/optical spectroscopy, although
for SMMJ09431+4700 the optical redshift is inferred from a source that
is physically distinct from the submm source. Our data more
than double the number of published, mm-confirmed SCUBA redshifts. In
addition to the three galaxies discussed here, these are SMMJ14011+0252
($z$=2.56; Frayer et al.\ 1999; Ivison et al.\ 2001; Downes \&
Solomon 2003) and SMMJ02399$-$0136 ($z$=2.81; Frayer et al.\ 1998;
Ivison et al.\ 1998; Genzel et al.\ 2003).  Our observations confirm
that at least some SCUBA galaxies are luminous and gas-rich systems
seen at a similar epoch to the UV-bright QSO and Ly-break galaxies
populations (Boyle et al.\ 2000; Steidel et al.\ 1999).\\
\indent All five SCUBA galaxies are rich in molecular gas. For the CO
luminosity to gas mass conversion factor appropriate for local ULIRGs
the median gas mass of the five SCUBA sources is 2.1 ($\pm$1.7)$\times
10^{10}$ M$_{\odot}$, similar to the median molecular gas masses found
in high-$z$ QSOs (e.g.\ Alloin et al.\ 1997; Downes et al.\ 1999;
Guilloteau et al.\ 1999; Barvainis et al.\ 2002; Cox et al.\ 2002) but
about three times greater than those of local ULIRGs (Solomon et al.\
1997). Assuming the most probable value for sin $i$=$2/\pi$, the
median ratio of gas mass to dynamical mass in the five galaxies is
$\sim$0.5, again 3 times greater than in ULIRGs (Downes \& Solomon
1998), and similar to the $z$=2.72 Ly-break galaxy cB58 (Baker et al.\
2003). Four of the five systems are composite AGN/starburst galaxies
in a complex environment, such as a merger/interacting system. The
fact that the submm galaxies are complex systems is the more
noteworthy as their redshift range is close to the peak of the merging
assembly history of galaxy evolution. Perhaps the multiple nature of
the SCUBA sources, along with the action of winds and outflows may
explain how Chapman et al.\ (2003) were able to see strong UV line
emission in sources as rich in gas and dust as the submm-population.\\
\indent Relative to their gas reservoir, submm galaxies are
very efficient emitters of radiation. The ratio of IR luminosity to
gas mass in our five sources has a median of 380 $\pm $ 170\thinspace
L$_{\odot }/$M$_{\odot }$, similar to high-$z$ QSOs (750 $\pm $ 350),
HzRGs (260 $\pm $ 70; Papadopoulos et al.\ 2000) and local ULIRGs (260
$\pm $ 160; Solomon et al.\ 1997), but significantly larger than local
LIRGs and more moderate luminosity starbursts (45 $\pm $ 30; Solomon
et al.\ 1997). A young starburst with a Salpeter IMF between 1 and 100
M$_{\odot } $ has $\sim$ 10$^{3}$ L$_{\odot}/$M$_{\odot}$. If
most of the IR luminosity of our submm sources is due to star
formation, their star formation efficiency must be high, or the IMF
must be biased toward high mass stars.\\
\indent Our observations strengthen the conclusion (e.g.\ Genzel et al.\
2003) that the brightest submm galaxies (S$_{850}\sim 2$--10
mJy, corrected for lensing) have dynamical masses within the central
few kpc that are comparable to massive, local early type galaxies
(M$_{\mathrm{gas}}$+M$_{ \mathrm{stars}}\sim $ m$^{\ast }\sim $
$7\times 10^{10}$\thinspace M$_{\odot }$; Cole et al.\
2001). Keeping in mind that our observations strictly give only upper
limits or rough estimates of source sizes, we obtain a median
dynamical mass of 5.5$\times 10^{10}$ M$_{\odot }\sim $ 0.8 m$^{\ast
}$, for $\sin^{-2}$ $i$=$\pi^{2}/4$. Current semianalytic models of
star formation in hierarchical CDM cosmogonies have difficulties
accounting for the observed space density of such massive baryonic
systems at $z$\,$\sim $\,3. These models predict too few m$^{\ast }$
galaxies at that redshift, by about an order of magnitude, perhaps as
a result of too slow baryonic cooling and low star formation
efficiencies in the models (Genzel et al.\ 2003).

\acknowledgements We are grateful to Prof.\thinspace M.Grewing for
granting us the discretionary time that made this project possible on
a short time scale. We also thank L.Tacconi for help with the data
reduction, S.Seitz for discussions on the lensing model for
SMMJ04431+0210, and I.Smail, A.Baker and D.Lutz for
thoughtful comments. We also thank O.Almaini and C.Willott for
permission to use their optical images of N2 850.4. AWB acknowledges
partial funding support by NSF grant AST0205937, and TRG from the EU
RTN Network POE.

\clearpage 
\begin{deluxetable}{lccccccc} 
\tabletypesize{\scriptsize}
\tablewidth{0pt}
\tablecaption{Observed Properties for the Three SMM Sources. \label{tbl-1}}
\tablehead{
\colhead{} & \colhead{} & \colhead{} & \colhead{R.A.} & \colhead{Decl.} &
\colhead{$I_{\rm CO}$}  & \colhead{Flux} & \colhead{Line Width} \\
\colhead{Source} & \colhead{Transition} & \colhead{Redshift} & \colhead{(J2000.0)} &
\colhead{(J2000.0)}  & \colhead{(Jy\,km\,s$^{-1}$)} & \colhead{(mJy)} & \colhead{(km\,s$^{-1}$)}}
\startdata
SMM J04431+0210 \hfill{\ldots} & CO(3--2) & 2.5094 $\pm$ .0002 & 04 43 07.25 & 02 10 23.3 & 1.4 $\pm$ 0.2$^{(a,c)}$ & \ldots & 350 $\pm$ 60 \\
                               & CO(7--6) & \nodata & \nodata  & \nodata & $\le 0.8 $ & \nodata & \nodata \\
                               & 3.0\,mm & \nodata & \nodata  & \nodata & & $\le 0.3$ &  \\
                               & 1.3\,mm & \nodata & \nodata  & \nodata & & 1.1 $\pm$ 0.3 &  \\
SMM J09431+4700 \hfill{\nodata} & CO(4--3) & 3.3460 $\pm$ .0001 & 09 43 03.74 & 47 00 15.3 & 1.1 $\pm$ 0.1$^{(a,b)}$ & \nodata & 420 $\pm$ 50 \\
                               & CO(9--8) & \nodata & \nodata  & \nodata & $\le 1 $ & \nodata & \nodata \\
                               & 2.8\,mm & \nodata & \nodata  & \nodata & & $\le 0.4 $ & \\
                               & 1.3\,mm & \nodata & 09 43 03.69 & 47 00 15.5 & & 2.3 $\pm$ 0.4 & \\
SMM J16368+4057 \hfill{\nodata} & CO(3--2) & 2.3853 $\pm$ .0014 & 16 36 50.43 & 40 57 34.7 & 2.3 $\pm$ 0.2$^{(a,b)}$ & \nodata & 840 $\pm$ 110 \\
                               & CO(7--6) & $\sim$2.383  & 16 36 50.41 & 40 57 34.3 & 1.1 $\pm$ 0.2& \nodata & \nodata \\
                               & 2.9\,mm & \nodata & \nodata  & \nodata & & $\le 1 $ & \\
                               & 1.3\,mm & \nodata & 16 36 50.40  & 40 57 34.2 & & 2.5 $\pm$ 0.4 & 
\enddata

\tablecomments{The astrometric accuracy is $\le 0.3''(=\sigma)$;
Limits on $I_{\rm CO}$ and Flux are 2\,$\sigma$, and uncertainties
include statistical errors, as well as absolute flux errors; Line
Width is full width at half maximum (FWHM); Line Velocity and
1\,$\sigma$ error rounded to 10 km\,s$^{-1}$; $^{a}$ No continuum
subtracted; $^{b}$ Gaussian fit; $^{c}$ Frayer et al.\ (2003) report
an upper limit of 2.5 Jy\,km\,s$^{-1}$ using the OVRO interferometer.}

\end{deluxetable}

\clearpage 
\begin{deluxetable}{lcccccc}
\tabletypesize{\footnotesize}
\tablecaption{Derived Properties for the Three SMM Sources. \label{tbl-2}}
\tablewidth{0pt}
\tablehead{
\colhead{} & \colhead{$D_A$} & \colhead{$A = \mu_L$\tablenotemark{e}} & \colhead{$D = 1''$\tablenotemark{a}} & 
\colhead{${{L'}_{\rm CO}}$\tablenotemark{a}} & \colhead{$M_{\rm gas}$\tablenotemark{a,d}} & \colhead{$L_{\rm FIR}$\tablenotemark{a,f}}\\
\colhead{Source} & \colhead{(Gpc)} & \colhead{} & \colhead{(Kpc)} & \colhead{($10^{10}$\,K\,km\,s$^{-1}$\,pc$^{-1}$)} &
\colhead{($10^{10}\,M_\odot$)} & \colhead{($10^{13}\,L_\odot$)}}
 \startdata
SMM J04431+0210 \hfill{\ldots} & 1.66 & 
%20.5 & 
4.4\tablenotemark{b} & 1.8 & 1.0 $\pm$ 0.2 & 0.8 & 0.3 \\
SMM J09431+4700 \hfill{\ldots} & 1.53 & 
%29.0 & 
1.2\tablenotemark{c} & 6.2 & 2.7 $\pm$ 0.3 & 2.2 & 1.7 \\
SMM J16368+4057 \hfill{\ldots} & 1.68 &
% 19.3 & 
\nodata & 8.1 & 6.9 $\pm$ 0.6 & 5.5 & 1.6 \enddata
\tablecomments{Adopting a flat cosmology of $\Omega_{\rm M} = 0.3$,
$\Omega_\Lambda = 0.7$ and $H=70$ km\,s$^{-1}$\,Mpc$^{-1}$; $^{a}$
Corrected for the lensing magnification $\mu_L$, if applicable; $^{b}$
From Smail et al.\ 1999; $^{c}$ From Cowie et al.\ 2002; $^{d}$
Adopting a conversion factor $\alpha = M_{\rm gas}/{L'}_{\rm CO}$ =
0.8 $M_{\odot}$\,(K\,km\,s$^{-1}$\,pc$^2)^{-1}$. ${L'}_{\rm CO}$ is
the apparent CO line luminosity corrected for the lensing
magnification (Solomon et al.\ 1997); $^{e}$ Assuming equal flux
amplification and linear magnification; $^{f}$ Obtained from 850$\mu$m
flux densities %through equ.\ 2 with a modified greybody model
% (T=40\,K) and a frequency dependent emissivity
(Blain et al.\ 2002).}
\end{deluxetable}

\clearpage 
\begin{figure}[tbp]
\vspace*{4cm}{\hspace*{-17.7cm}{\ 
\plotfiddle{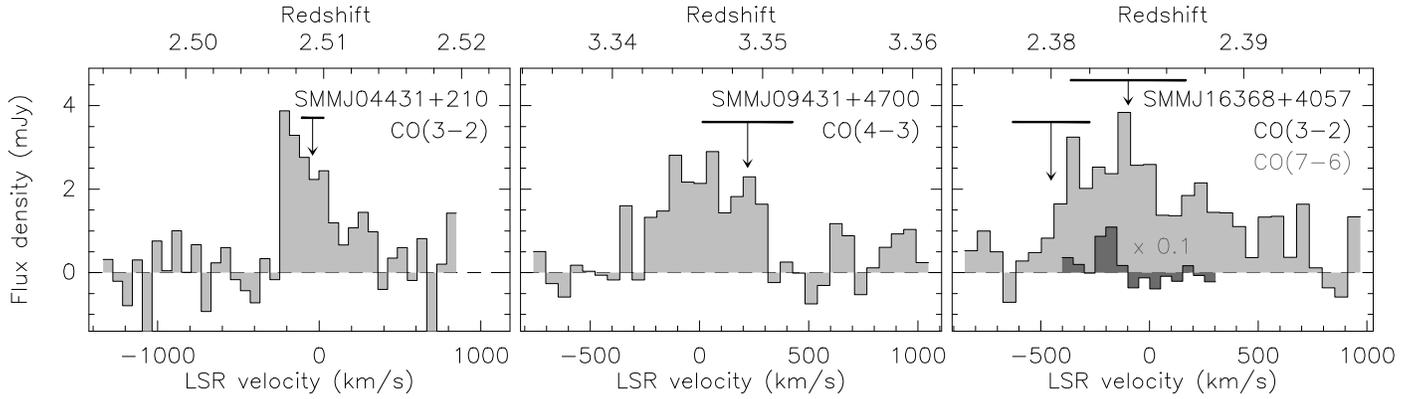}{0cm}{-90}{.75}{!}{0cm}{0cm}}}\vspace*{-5cm}
\caption{CO spectra of the three SCUBA sources. The LSR velocity scale
is with respect to the CO redshift listed in Table 1. Optical
redshifts (arrows, horizontal bars are uncertainties) are from Frayer
et al.\ 2003 (H$\protect\alpha$+[N\,\textsc{ii}], left panel), Ledlow
et al.\ 2002 (Ly$\protect\alpha$, center) and Smail et al.\ 2003 (UV
photospheric + Seyfert 2), respectively. The rms noise per 20 MHz
channel is 0.7, 0.9 and 0.6 mJy for the three sources, respectively
(from left to right), in the spectral region where the frequency
settings were overlapping, and increases to about 20 percent toward
the edges of the bandpass. Overplotted on the CO 3-2 spectrum of
SMMJ16368+4057 is the CO 7-6 spectrum scaled down to one-tenth of its
flux density. The rms noise per 40 MHz channel is here 2.2 mJy.}
\end{figure}

\clearpage

\begin{figure}[tbp]
\vspace*{4cm}{\hspace*{-17.7cm}{\ 
\plotfiddle{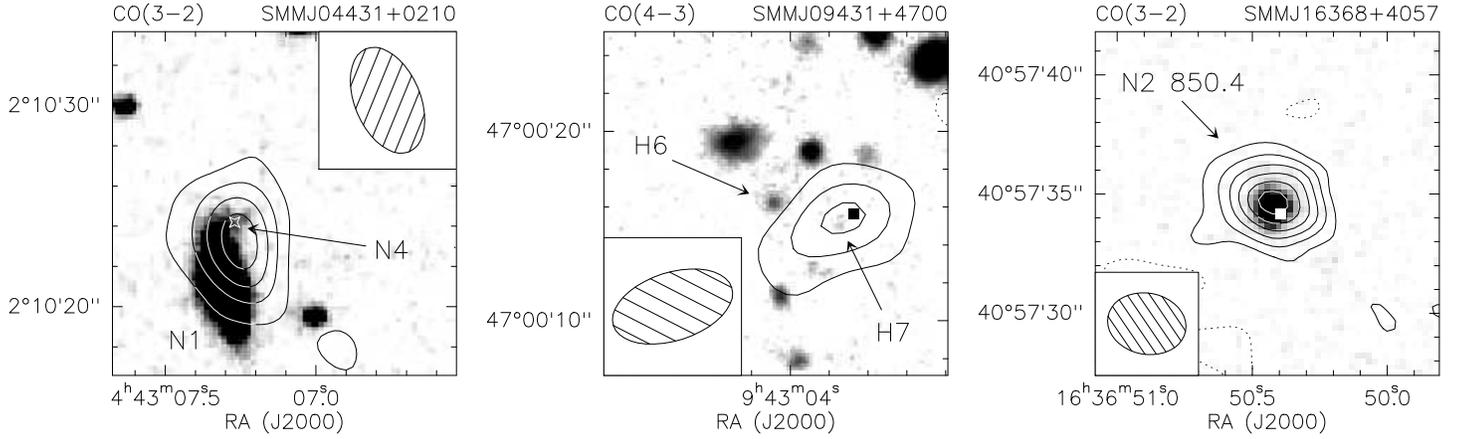}{0cm}{-90}{.7}{!}{0cm}{0cm}}}\vspace*{-3cm}
\caption{Velocity integrated, natural weighted CO maps of the three
SCUBA sources, superposed on greyscale images of the optical
emission. The contours are in units of 2$\protect\sigma$ of the noise
level, and are 0.26, 0.31 and 0.20 Jy\,km\,s$^{-1}$ in the three
panels (left to right), respectively. The synthesized beams are (left
to right, shown as hatched ellipses) 5.6$\times$3.3$ ^{\prime\prime}$
at position angle 23$^{\circ}$ (E of N), 6.6$\times$
3.6$^{\prime\prime}$ at 108$^{\circ}$, and 3.3$\times$2.6$
^{\prime\prime}$ at 79$^{\circ}$. The three underlying images are in
the $K$-band (left panel: Frayer et al.\ 2003, and right panel: Smail
et al.\ (2003) and in the $I$-band (center panel: Ledlow et al.\
2002). The asterisk (left image) is the position of the ERO N4
(uncertainty $\pm 0.5^{\prime\prime}$; see text), the filled squares
(black and light) are the millimeter continuum positions (center and
right).  The edge on spiral galaxy 2$^{\prime\prime}$ SE of the CO
source is the source N1 in the foreground cluster at redshift
$z$=0.18. In the middle panel the positions of the two radio sources
H6 and H7 are denoted by arrows.  The stronger optical and radio
source H6 is a narrow line Sey1 galaxy at redshift $z$=3.349
(Ly$\protect\alpha$). The CO 4-3 emission from N2 850.4 remains
largely unresolved.}
\end{figure}
\end{document}